\documentclass[11pt,preprint]{aastex}      

\usepackage[latin1]{inputenc}
\usepackage{lscape}
\usepackage{amsmath}

\slugcomment{\footnotesize{Accepted for publication in ApJ Letters (2012)}}

\shortauthors{Carrasco-Gonz\'alez et al.}

\begin{document}

\title{Resolving the Circumstellar Disk Around the Massive Protostar Driving the HH 80-81 Jet}

\author{Carlos~Carrasco-Gonz\'alez\altaffilmark{1},
Roberto~Galv\'an-Madrid\altaffilmark{2},
Guillem~Anglada\altaffilmark{3},
Mayra~Osorio\altaffilmark{3},
Paola~D'Alessio\altaffilmark{4},
Peter Hofner\altaffilmark{5,6},
Luis~F.~Rodr\'{\i}guez\altaffilmark{4},
Hendrik Linz\altaffilmark{7},
Esteban D. Araya\altaffilmark{8}}

\altaffiltext{1}{Max-Planck-Institut f\"ur Radioastronomie (MPIfR), Auf dem H\"ugel 69, 53121, Bonn, Germany; carrasco@mpifr-bonn.mpg.de}
\altaffiltext{2}{European Southern Observatory, Karl-Schwarzschild-Str. 2, 85748 Garching, Germany}
\altaffiltext{3}{Instituto de Astrof\'{\i}sica de Andaluc\'{\i}a, CSIC, Camino Bajo de Hu\'etor 50, E-18008 Granada, Spain}
\altaffiltext{4}{Centro de Radioastronom\'{\i}a y Astrof\'{\i}sica UNAM, Apartado Postal 3-72 (Xangari), 58089 Morelia, Michoac\'an, Mexico}
\altaffiltext{5}{Physics Department, New Mexico Tech, 801 Leroy Pl., Socorro, NM 87801, USA}
\altaffiltext{6}{National Radio Astronomy Observatory, P.O. Box O, Socorro, NM 87801, USA}
\altaffiltext{7}{Max-Planck-Institut f\"ur Astronomie (MPIA)}
\altaffiltext{8}{Physics Department, Western Illinois University, 1 University Circle, Macomb, IL 61455, USA}

\begin{abstract}

We present new high-angular resolution observations toward the driving source of the HH~80--81 jet (IRAS 18162--2048). Continuum emission was observed with the Very Large Array at 7 mm and 1.3 cm, and with the Submillimeter Array at $860~\mu$m, with angular resolutions of $\sim0\farcs1$ and $\sim0\farcs8$ respectively. Submillimeter observations of the sulfur oxide (SO) molecule are reported as well. At 1.3 cm the emission traces the well-known radio jet, while at 7 mm the continuum morphology is quadrupolar and seems to be produced by a combination of free-free and dust emission. An elongated structure perpendicular to the jet remains in the 7 mm image after subtraction of the free-free contribution. This structure is interpreted as a compact accretion disk of $\sim200$ AU radius. Our interpretation is favored by the presence of rotation in our SO observations observed at larger scales. The observations presented here add to the small list of cases where the hundred-AU scale emission from a circumstellar disk around a massive protostar has been resolved. 

\end{abstract}


\section{Introduction}
 
It is well known that the formation of solar-type stars takes place with the assistance of an accretion disk that transports gas and dust from the envelope of the system to the protostar, and a jet that removes angular momentum from the system, allowing accretion to proceed (e.g., McKee \& Ostriker 2007). While it is tempting to think that these mechanisms work all the way up in the stellar-mass range, it is not clear to what extent this assertion may be correct. Feedback from a growing protostar increases rapidly with mass, and early calculations suggested that it could be fatally disruptive for stellar masses $M_\star \ga 8~M_\odot$ (Wolfire \& Cassinelli 1987). It has been proposed that stars of higher mass are able to form via accretion when revised dust opacities and high mass-accretion rates are considered (Osorio et al. 1999). However, it is still unclear if processes such as radiation pressure, ionizing radiation, and jet/outflow feedback can terminate accretion onto the most massive young stellar objects, and if they do, how and when does this happen (for a review see, e.g., Zinnecker \& Yorke 2007). 

Theory suggests that several mechanisms can be at work to aid accretion onto massive young stellar objects (MYSOs). Flattened accretion flows focus the disruptive effects of radiation pressure (e.g., Yorke \& Sonnhalter 2002, Kuiper et al. 2011), photoionization (e.g., Peters et al. 2010), and jets and outflows (e.g., Wang et al. 2010, Cunningham et al. 2011) to some preferential angles while permitting accretion from other directions. Observations support this view; there are several known cases of flattened, dense gas structures that appear to be rotating and infalling around MYSOs (e.g., Cesaroni et al. 2007). However, these structures usually do not look like the Keplerian, stable disks with sizes of $\la$100~AU seen around solar-type YSOs. Instead, they often appear very large, up to $\sim 10^4$ AU in diameter, and very massive compared to their central star(s), which therefore renders them unstable to fragmentation. In addition, they show apparent infall motions that are comparable in magnitude to their rotation. Indeed, model fitting of the SEDs of a sample of massive protostar candidates (De Buizer et al. 2005) suggests that the observed large-scale structures with sizes of thousands of AU could be naturally explained as infalling flattened envelopes, while the formation of the "true" accretion disks is expected to occur at scales of the order of the centrifugal radius (a few hundred of AU for these MYSOs).

A more direct approach to test the presence of circumstellar disks in MYSOs is to look for "clean" examples of systems composed of a compact disk and a jet. These cases appear to be quite rare. Examples are G192.16--3.82 (Shepherd et al. 2001), AFGL 2591 (Trinidad et al. 2003), IRAS~18162--2048 (G\'omez et al. 2003), Cepheus A HW2 (Patel et al. 2005), IRAS~20126+4104 (Hofner et al. 2007), IRAS~16547--4247 (Franco-Hern\'andez et al. 2009), and IRAS~13481--6124 (Kraus et al. 2010). However, up to now, only Cepheus~A~HW2 has been observed with enough angular resolution and sensitivity to angularly resolve the emission of the disk; in the other cases, the disk emission has not been well resolved, or evidence for the accompanying jet is weak. 

One of the best candidates to look for a circumstellar disk is the massive protostar IRAS 18162--2048. This protostar has a bolometric luminosity of $L\sim$2$\times$10$^4$~L$_\sun$ (Aspin \& Geballe 1992), equivalent to that of a B0 zero-age main sequence star ($M_* \ga 10~M_\sun$). It powers a highly-collimated radio jet that extends 5.3~pc (at an adopted distance of 1.7 kpc; Rodr\'{\i}guez et al. 1980) toward the Herbig-Haro objects HH 80--81--81N (Mart\'{\i} et al. 1993, 1998). The jet is surrounded by a bipolar cavity seen at 8 $\mu$m (Qiu et al. 2008). Furthermore, it has been found that this jet is being collimated by a large-scale helical magnetic field, most probably originated in a rotating accretion disk (Carrasco-Gonz\'alez et al. 2010). G\'omez et al. (2003) reported unresolved observations of the mm thermal dust emission from the exciting source of the jet. Recently, Fern\'andez-L\'opez et al. (2011a,b) presented (sub)mm observations down to an angular resolution $\sim 0\farcs5$ ($\sim$850~AU), and interpreted the emission as arising from a compact (size $\lesssim$ 600~AU) accretion disk orbiting a $\sim 15~M_\odot$ central source.

In this Letter, we present new sensitive observations performed with the Submillimeter Array (SMA) and the Very Large Array (VLA) towards the IRAS 18162--2048 MYSO. These observations resolve, for the first time, the dust emission of this utmost important source at angular resolutions down to $\sim 0\farcs1$, equivalent to $\sim 170$ AU. 

\section{Observations}

\subsection{SMA Observations}
 
Observations in the 0.8-mm band were performed with the SMA\footnote{The Submillimeter Array is a joint project between the Smithsonian Astrophysical Observatory and the Academia Sinica Institute of Astronomy and Astrophysics and is funded by the Smithsonian Institution and the Academia Sinica.}  (Ho et al. 2004) during two runs (2006 June 13 and 22). In the first run, the array was in its extended configuration, while in the second run, the array was in the compact configuration. Two sidebands spanning the frequency ranges 342.6-344.6 GHz and 352.6-354.6 GHz were covered. Calibration was performed using the MIR data calibration program. Quasars 3C454.3 and J1924--292 served as bandpass and phase calibrators, respectively. The absolute flux scale was derived from observations of Callisto, and is accurate to better than $\sim 15\%$. Further processing and imaging was done in MIRIAD, AIPS, and IDL. In addition to the continuum emission, we report on the detection of compact emission of SO~8(8)--7(7) ($\nu_0$=344.31061 GHz) around the core of the radio jet. HCN (4-3) and CS (7-6) were also detected but their emission is not confined to the immediate surrounding of the exciting source and their analysis is out of the scope of this Letter.
 
The final continuum map was done from the extended configuration data with a uniform weighting to maximize the angular resolution (synthesized beam~=~0$\farcs$97$\times$0$\farcs$70, P.A.= $-$26$^\circ$). The SO map was made from the concatenated compact+extended data with an intermediate weighting (robust=0) to have the best compromise between resolution and sensitivity (synthesized beam~=~1$\farcs$20$\times$0$\farcs$99, P.A.= $-$34$^\circ$). 

\subsection{VLA Observations}

Observations at 1.3~cm and 7~mm continuum were carried out using the VLA of the National Radio Astronomy Observatory (NRAO)\footnote{The NRAO is a facility of the National Science Foundation operated under cooperative agreement by  Associated Universities, Inc.}\ in its A configuration on 2004 November 7 (1.3 cm and 7 mm) and 20 (1.3 cm). Phase and flux calibrators were J1820$-$254 and 3C286, respectively. Data editing and calibration were performed using the AIPS package, following the standard high-frequency VLA procedures. 
 
Maps at 1.3~cm and 7~mm were made applying a tapering of 1750 k$\lambda$ and 2100 k$\lambda$, respectively, in order to emphasize extended emission. Synthesized beams are 0$\farcs$19$\times$0$\farcs$13 with a position angle (P.A.) of 20$^\circ$ (1.3 cm) and 0$\farcs$12$\times$0$\farcs$09 with a P.A. of 26$^\circ$ (7 mm).
   
In order to compare the SMA observations with the emission of the HH 80--81 radio jet at similar scales, we calibrated VLA A configuration archive data at 3.6 cm continuum obtained in 5 epochs (1990.2, 1994.3, 1995.5, 1997.1 and 2006.4). The map shown in this paper was made by concatenating data from all the epochs (synthesized beam=0$\farcs$5$\times$0$\farcs$3; P.A.=0$^\circ$).

\section{Results and Discussion}

\subsection{An infalling rotating molecular envelope}

\begin{figure}  
\begin{center}

\includegraphics[width=\textwidth]{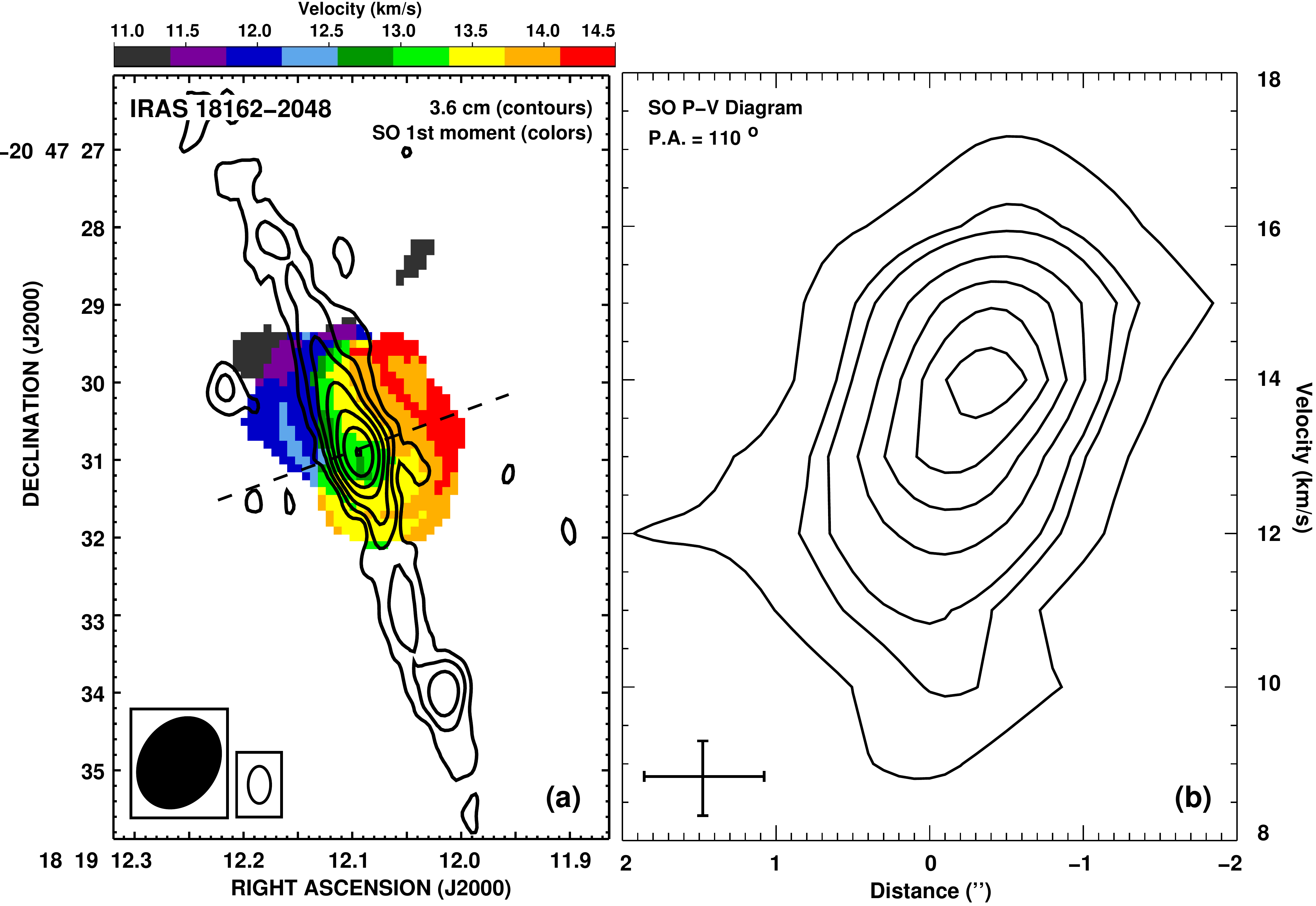}  

\caption{\footnotesize{\textbf{(a)} Superposition of the 3.6~cm VLA contour map over the SMA first-order moment (velocity) color map of the SO molecule emission. Contours are $-$3, 3, 6, 12, 25, 50, 100, and 200 times the rms of the VLA map, 10 $\mu$Jy~beam$^ {-1}$. Color scale ranges from 11.0 to 14.5 km~s$^{-1}$. \textbf{(b)} Position-Velocity diagram at a P.A. of 110$^\circ$ (perpendicular to the radio jet) centered on the core of the radio jet. Contour levels are 0.4, 0.8, 1.0, 1.4, 1.8, 2.2 and 2.6 Jy~km~s$^{-1}$~beam$^{-1}$. Spatial and velocity resolutions are represented by the cross in the lower-left corner.}}
\label{Fig1}  
\end{center} 
\end{figure}

In Figure \ref{Fig1}a we show a superposition of the VLA map at 3.6~cm (contours) over the first moment of the SO molecule emission (colors) obtained with the SMA. At 3.6~cm, the radio jet appears with an elongated morphology along a P.A. of 20$^\circ$. The SO molecule emission shows an extended envelope (size$\simeq$3000 AU) around the driving source of the radio jet, with a velocity gradient roughly perpendicular to it (see Fig. \ref{Fig1}a), that we interpret as rotational motions. In Figure \ref{Fig1}b, we show a position-velocity diagram along a direction perpendicular to the jet. From this diagram we measure a velocity gradient of $\sim2.5$~km~s$^{-1}$ arcsec$^{-1}$, from which we infer a rotation velocity $\sim$2 km~s$^{-1}$ at a radius $\sim$ 1500 AU, assuming an inclination angle of 90$^\circ$ (i.e., the HH~80-81 jet is almost in the plane of the sky). The centrifugal radius (i.e., the largest radius on the equatorial plane that receives the infalling material) is given by $R_c=r_0^2 v_0^2/(GM)$, where $v_0$ is the rotation velocity at a distant reference radius $r_0$, and $M$ is the central mass\footnote{This expression for $R_c$ is derived assuming conservation of the specific angular momentum. Angular momentum losses during the infall process would decrease the actual value of the centrifugal radius.}. Adopting the values of $r_0$ and $v_0$ derived from our SO observations, and assuming $M \simeq 15~M_\odot$ (Fern\'andez-L\'opez et al. 2011b), we obtain $R_c \simeq 650$ AU. This result strongly suggests that the SO emission, which arises from radii larger than $R_c$, is tracing an infalling and rotating envelope, while the "true" accretion disk should be formed at smaller scales, within the centrifugal radius.

\subsection{A compact dusty disk}

\begin{figure}
\begin{center}

\includegraphics[width=\textwidth]{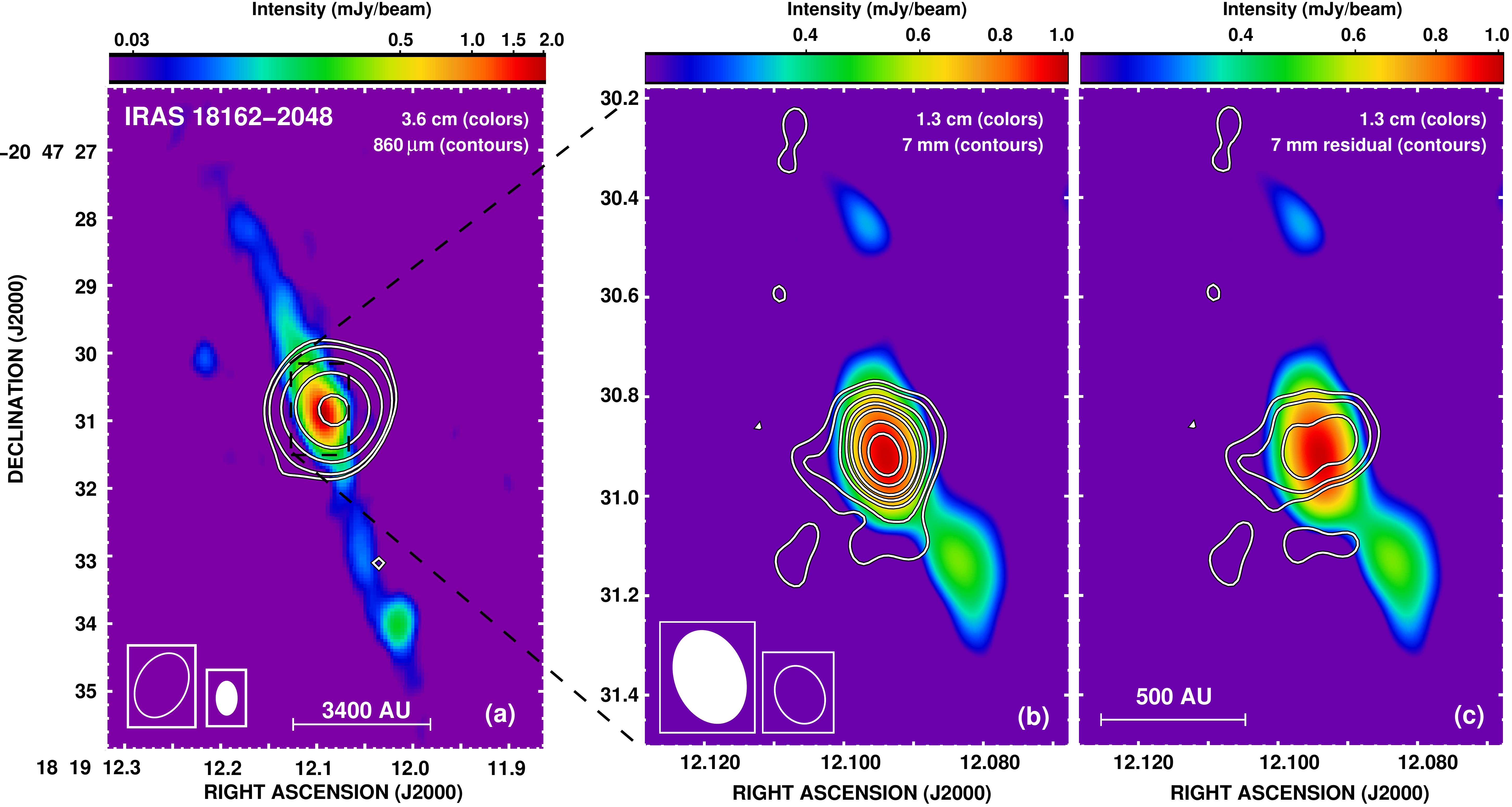}  

\caption{\footnotesize{\textbf{(a)} Superposition of the 860~$\mu$m continuum SMA extended configuration map (contours) over the 3.6 cm VLA map (colors). Contours are $-$3, 3, 4, 8, 16, and 32 times the rms of the SMA map, 10.7 mJy~beam$^{-1}$. \textbf{(b)} Superposition of the 7~mm VLA map over the 1.3~cm VLA map. Contour levels are 3, 4, 6, 8, 10, 14 and 18 times the rms of the map, 120 $\mu$Jy~beam$^ {-1}$. \textbf{(c)} Same as (b), after subtraction of the free-free contribution at 7~mm.}} 

\label{Fig2}  
\end{center} 
\end{figure}

In Figure \ref{Fig2}a we show a superposition of the 860~$\mu$m continuum map (contours) over the 3.6~cm map of the radio jet (colors). The compact 860~$\mu$m continuum emission is observed towards the core of the radio jet and has a flux density of 580$\pm$10~mJy. A Gaussian fit to the 860~$\mu$m source gives a deconvolved FWHM $\lesssim$0$\farcs$7 ($\la$1200 AU). Given that the spectral index $\alpha$ (where $S_\nu \propto \nu^\alpha$ ) of the free-free jet is $\sim 0.2$ at cm wavelengths (Mart\'{\i} et al. 1993), the free-free contribution at 860~$\mu$m should be $\lesssim 3$ mJy ($0.5~\%$). Therefore, the submm emission is dominated by dust, likely from an accretion disk (see below), but it remains unresolved at the SMA angular resolution of $\sim$0$\farcs$8 ($\sim$1400 AU).  

In Figure \ref{Fig2}b we show the superposition of our VLA 7~mm (contours) and 1.3~cm (color scale) maps covering the central region of the radio jet. The very high angular resolution of these maps ($\sim$0$\farcs$15, equivalent to 250 AU at 1.3~cm; $\sim$0$\farcs$10, equivalent to 170 AU at 7~mm) allows us to resolve the structure of the core of the source. The 1.3~cm emission shows a jet-like morphology, consisting of a bright central source and two weaker sources to the NE and SW. The global orientation of the 1.3 cm emission is similar to that of the larger scale radio jet detected with a higher signal-to-noise ratio at 3.6~cm. However, the central 1.3~cm source is oriented at P.A.$\simeq$10$^\circ$ at scales of $\sim$0$\farcs$1-0$\farcs$2, while at larger scales ($\sim$0$\farcs$5) the jet is oriented at P.A.$\simeq$20$^\circ$. This suggests a precession of the jet axis, as previously proposed by Mart\'{\i} et al. (1993) for the large-scale (10$'$) jet. 

\begin{figure}  
\begin{center}

\includegraphics[width=\textwidth]{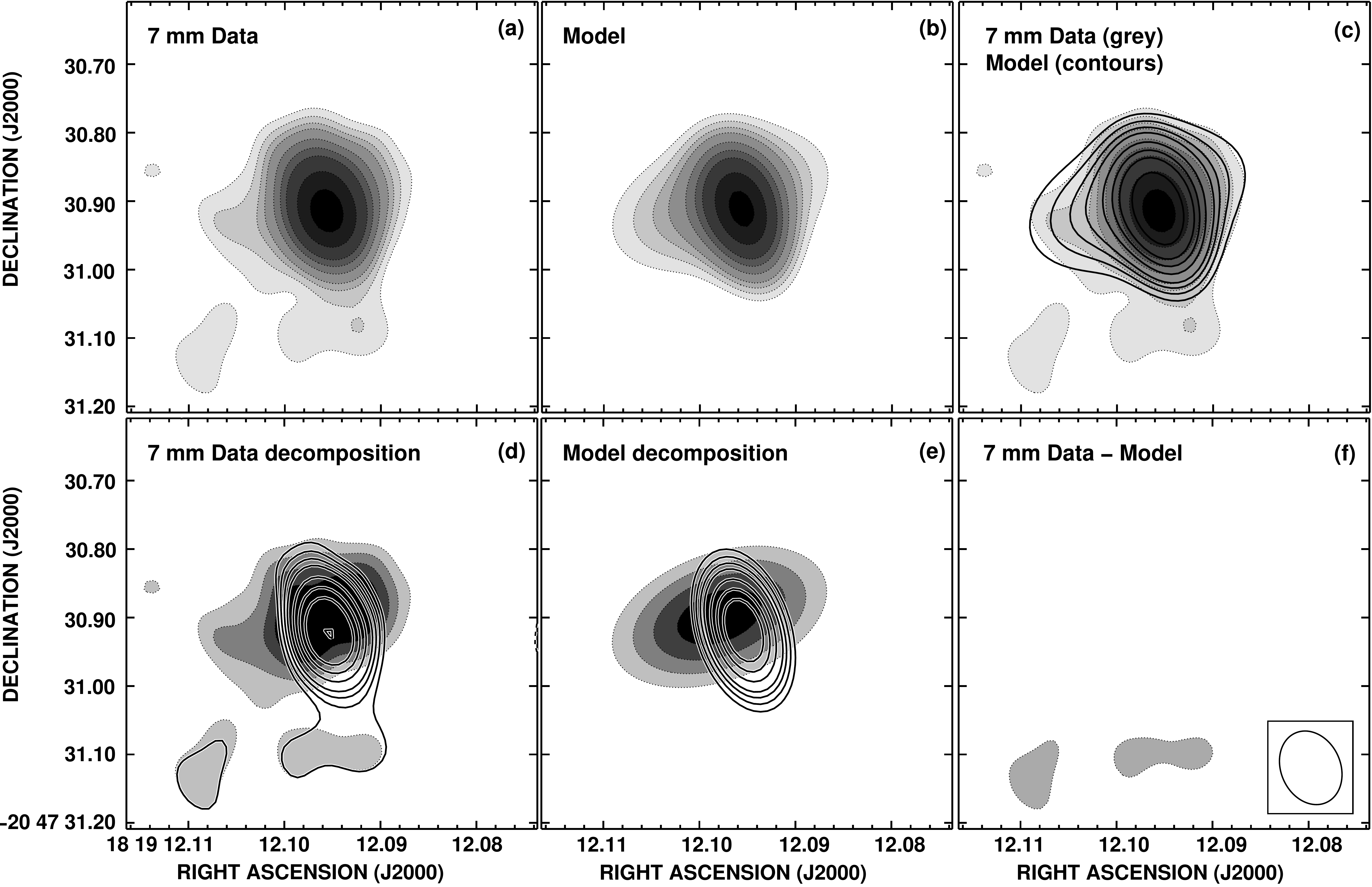}  

\caption{\footnotesize{Decomposition of the quadrupolar 7 mm source in two perpendicular Gaussian sources. Panels (a) and (b) show the VLA 7 mm map and the fitted model, respectively. Panel (c) shows a comparison of the 7 mm data and the model. Panel (d) shows the decomposition of the 7 mm emission in two perpendicular components. The N-S (respectively, E-W) emission is obtained by substracting to the 7 mm data the E-W (respectively, N-S) model. Panel (e) shows the two Gaussian model sources whose parameters are given in Table \ref{Tab1}. We note a small difference in the central positions of the components that, if real, could be due to density inhomogeneities in the jet and/or the disk. Panel (f) shows the residual after subtracting the model to the 7 mm data. In all panels, contours are $-$3, 3, 4, 5, 6, 8, 10, 12, 16, and 20 times the rms noise of the 7 mm map, 120 $\mu$Jy~beam$^{-1}$. }}

\label{Fig3}  
\end{center} 
\end{figure}

Emission at 7~mm is detected only towards the center of the radio jet and shows a quadrupolar morphology which can be described as the superposition of two overlapping, elongated sources. We have fitted the 7~mm source with two Gaussian ellipsoids. This fit is shown in Figure \ref{Fig3} and the obtained parameters are listed in Table \ref{Tab1}. Both components are extended with perpendicular orientations (roughly N-S and E-W) and similar flux densities ($\sim$3~mJy; see Table \ref{Tab1}).   

Similar quadrupolar morphologies have been observed in other radio jets of low-mass stars at wavelengths where comparable contributions of free-free emission from the radio jet and thermal dust emission from a perpendicular disk are present (e.g., HH~111 at 7~mm: Rodr\'{\i}guez et al. 2008; HL~Tau at 1.3~cm: Carrasco-Gonz\'alez et al. 2009). The total flux density at 7~mm ($\sim$6.3 mJy) and the flux density of the central 1.3 cm component (see Table \ref{Tab1}) imply a spectral index $\alpha \simeq$2, much higher than the spectral index of the ionized jet, strongly suggesting the presence of thermal dust emission at 7~mm. Therefore, we interpret the morphology of the 7~mm emission as the result of a combination of free-free emission from an ionized jet (N-S component) and thermal dust emission from a perpendicular structure (E-W component). In Figure \ref{Fig2}c we show a superposition of the E-W component (after substraction of the free-free contribution) over the 1.3 cm emission of the radio jet. From the deconvolved major axis of the E-W component (see Table \ref{Tab1}), we derive a radius of $\sim$200 AU, consistent with the radius ($\la$300~AU) of the unresolved 1.4 mm continuum source detected by Fern\'andez-L\'opez et al. (2011a) with the SMA. The radius obtained from our 7 mm observations is smaller than the centrifugal radius inferred from our SO observations ($R_c\simeq$ 650 AU). Thus, although our observations lack direct kinematical information, we interpret the 7 mm emission elongated in the E-W direction as tracing a dusty accretion disk. 
 
The inferred radius of the disk ($\sim$200 AU) is similar to that of the rotating disk around the massive protostar Cep A HW 2 ($\sim$300~AU; Patel et al. 2005), and fully consistent with the theoretical estimates for MYSOs (De Buizer et al. 2005). It is also of the same order (although somewhat larger) than typical values of disks around low-mass protostars ($\lesssim$100 AU, e.g., Andrews et al. 2009).

A rough estimate of the disk mass can be obtained from the dust emission at 7 mm using the equation

\begin{equation}
\left[ \frac{M_{\mathrm{disk}}}{\rm M_\sun} \right] = 0.16 \left[ \frac{\nu}{\mathrm{GHz}} \right]^{-2} \left[ \frac{\kappa_\nu}{\rm cm^2~g^{-1}} \right]^{-1} \left[ \frac{S_\nu}{\rm Jy} \right] \left[ \frac{T_d}{\rm K} \right]^{-1} \left[ \frac{D}{\mathrm{pc}} \right]^2,
\end{equation}
 
\noindent where $\kappa_\nu$ is the dust opacity per gram of dust+gas at frequency $\nu$, $S_\nu$ is the flux density, $T_d$ is the dust temperature, and $D$ is the distance to the region. The main sources of uncertainty in the mass determination are $T_d$ and $\kappa_\nu$ (sensitive to the distribution of grain sizes and to the presence of ice mantles on the grains). Since the upper-level energy of the observed SO transition is 87.45~K, then the temperature at 1500 AU scales should be $\ga$90 K. Assuming that the temperature varies with the radius as $T\propto R^{-0.5}$, we obtain a lower limit of $\ga$250~K for the temperature at 200 AU (the radius of the disk). For the 7 mm opacity we adopt a value of 0.001 cm$^2$ g$^{-1}$ (assumes a gas-to-dust ratio of 100), appropriate for warm ($T \sim 300$ K) gas with a grain maximum size of 1 mm (D'Alessio et al. 2001). With these assumptions and using Eq. 1, we estimate $M_{\rm disk} \la $4 M$_\sun$. This would imply a ratio of the disk to stellar mass $M_\mathrm{disk}/M_\star \la$0.3 which is towards the upper end of typical values found for low-mass protostars ($M_\mathrm{disk}/M_\star \sim$0.001--0.1; e.g., Andrews et al. 2009, Ricci et al. 2010). Our estimate for the disk mass is consistent with the value given by Fernandez-L\'opez et al. (2011a). However, it should be noted that both estimates are very uncertain since they have been obtained from very simple assumptions. In order to obtain an accurate estimate of the mass of the disk it would be necessary to model the SED over a wide range of frequencies using an accretion disk model that takes into account effects such as the geometry, physical properties, and heating from the protostar or the dusty envelope in a physically self-consistent mode.

The accretion rate of the disk onto the star cannot be easily calculated from our present data. However, a lower limit to the accretion rate can be estimated because this should be higher than the mass-loss rate in the jet. We assume a pure hydrogen jet with constant opening angle $\theta_0$, terminal velocity $v_\mathrm{jet}$, ionization fraction $x_0$, and electron temperature $T_e=10^4$ K. We further assume that the jet axis is in the plane of the sky. Under these assumptions, and following Reynolds (1986), the mass loss rate in the jet is given by

\begin{equation}
\left[ \frac{\dot{M}_\mathrm{out}}{\rm M_\sun~yr^{-1}} \right] = 1.9 \times 10^{-6} \, x_0^{-1} \, \left[ \frac{v_{jet}}{1000 \, \rm km~s^{-1}} \right] 
\left[ \frac{S_\nu}{\rm mJy} \right]^{0.75} \left[ \frac{\nu}{\rm GHz} \right]^{-0.45} \left[ \frac{D}{\rm kpc} \right]^{1.5} \left[ \frac{\theta_0}{\mathrm{rad}} \right]^{0.75}.
\end{equation}

The opening angle of the jet is estimated to be $\theta_0$=$2\arctan (\theta_\mathrm{min}/\theta_\mathrm{maj})$, where $\theta_\mathrm{min}$ and $\theta_\mathrm{maj}$ are the deconvolved minor and major axes of the jet, respectively. Using the deconvolved sizes and flux of the jet component from Table 1, taking $v_\mathrm{jet}$=1000 km~s$^{-1}$ (Mart\'{\i} et al. 1995), and assuming $x_0$=0.1 (e.g., Rodr\'{\i}guez et al. 1990, Shang et al. 2007), we estimate a mass-loss rate $\dot{M}_\mathrm{out}\sim$10$^{-5}$ M$_\sun$ yr$^{-1}$. The accretion rate $\dot{M}_\mathrm{acc}$ of the disk onto the star is expected to be $\sim$10 times larger than the mass-loss rate (e.g., Bontemps et al. 1996), or $\dot{M}_\mathrm{acc} \simeq 10^{-4}$ $M_\odot$~yr$^{-1}$. This value of the accretion rate is $\sim100$ times larger than typical values for low-mass protostars ($\dot{M}_\mathrm{acc} < 10^{-6}$ $M_\odot$~yr$^{-1}$; Evans et al. 2009). 

\section{Conclusions}

We found that the IRAS 18162-2048 massive protostar, that was known to be associated with the highly collimated jet HH~80--81, is surrounded by a compact disk that is angularly resolved in our observations. The observed radius of the disk ($\sim$ 200 AU, similar to the radius of the Cep~A~HW2 disk) is in agreement with the values theoretically expected for massive protostars, and somewhat larger than typical values observed in low-mass protostars. The ratio between the mass of the disk and that of the central star ($\sim$0.3) is within the range (although towards the upper end) of typical values found for low-mass stars. The mass accretion rate of the disk onto the star ($\sim$10$^{-4}$ M$_\sun$~yr$^{-1}$) seems to be much higher than in the low-mass case. These results reinforce the idea that the formation of high-mass stars is governed by physical mechanisms that are similar (but scaled-up) to those of low-mass star formation.

\emph{Acknowledgements.} We thank an anonymous referee for useful comments. This work was partially funded by the ERC Advanced Investigator Grant GLOSTAR (247078). R.G.-M. acknowledges funding from the European Community's Seventh Framework Programme (/FP7/2007-2013/) under grant agreement No 229517R. G.A., C.C.-G., and M.O. acknowledge support from MICINN (Spain) grants AYA2008-06189-C03-01 and AYA2011-30228-C03-01 (co-funded with FEDER funds) and from Junta de Andaluc\'{\i}a (TIC-126).  P.H. acknowledges support from NSF grant AST-0908901. L.F.R. and P.D. acknowledge the support of DGAPA, UNAM, and CONACyT (M\'exico).


\begin{deluxetable}{lcccc}
\tablewidth{0pt}
\tablecaption{Parameters of continuum sources\label{Tab1}}
\startdata
\hline \hline 
Wavelength	  & \multicolumn{2}{c}{Position (J2000)\tablenotemark{a}} &	 Flux Density  &   Deconvolved  			  \\ \cline{2-3}
		      &	      RA	       &	  DEC		                  &	    (mJy)      &   Angular Size\tablenotemark{b}	  \\ \hline
1.3 cm                      &   18 19 12.094    &  $-$20 47 30.91  &  1.7 $\pm$ 0.2  &  0$\farcs$23 $\times$ 0$\farcs$07 ;  10$^\circ$ $\pm$  7$^\circ$ \\
7 mm (N-S)\tablenotemark{c} &   18 19 12.094    &  $-$20 47 30.92  &  2.7 $\pm$ 0.3	 &  0$\farcs$12 $\times$ 0$\farcs$02 ;  11$^\circ$ $\pm$  7$^\circ$ \\
7 mm (E-W)\tablenotemark{c} &   18 19 12.096    &  $-$20 47 30.90  &  3.6 $\pm$ 0.5	 &  0$\farcs$23 $\times$ 0$\farcs$13 ; 110$^\circ$ $\pm$ 10$^\circ$ \\
860 $\mu$m                  &   18 19 12.084    &  $-$20 47 30.84  &  580 $\pm$ 10   &           $<$0$\farcs$7                                          \\
\enddata
\tablenotetext{a}{Units of right ascension are hours, minutes, and seconds and units of declination are degrees, arcminutes, and arcseconds. The absolute positional accuracy is estimated to be 0$\farcs$05.}
\tablenotetext{b}{Major axis $\times$ minor axis; position angle of major axis. Uncertainty in major and minor axis is estimated to be 0$\farcs$02.}

\tablenotetext{c}{Emission at 7~mm shows a quadrupolar morphology (see text). Parameters are obtained by fitting the 7 mm emission with two Gaussian ellipsoids using the task JMFIT. As initial guesses for the fit we used for both components the same position (that of the peak of the total 7 mm emission) and the same peak intensity (half of the peak of the total 7 mm emission). For the initial P.A.s, we used 10$^\circ$ (as inferred from the 1.3 cm source) and 100$^\circ$. Fits using different initial values converge to final values within the uncertainties.}

\end{deluxetable}




\end{document}